\documentstyle[12pt,epsfig,a4]{article}

\font\elevenrm=cmr10 scaled \magstep1

\def\sweff{\hbox{$\sin^2\theta_{eff}$}}

\def\rs{\hbox{$\sqrt s$}}

\def\Mcs{\hbox{MeV/$c^2$}}
\def\Gcs{\hbox{GeV/$c^2$}}

\def\epem{\hbox{$\hbox{e}^+\hbox{e}^-$}}
\def\e{\hbox{e}}

\def\ppb{\hbox{$\hbox{p}\overline{\hbox{p}}$}}
\def\p{\hbox{p}}
\def\gaga{\hbox{$\gamma\gamma$}}
\def\nunu{\hbox{$\nu\bar\nu$}}
\def\lplm{\hbox{$\ell^+\ell^-$}}

\def\toto{\hbox{$\tau^+\tau^-$}}

\def\qqb{\hbox{$\hbox{q}\overline{\hbox{q}}$}}
\def\bbb{\hbox{$\hbox{b}\overline{\hbox{b}}$}}
\def\ttb{\hbox{$\hbox{t}\overline{\hbox{t}}$}}

\def\d{\hbox{d}}
\def\u{\hbox{u}}
\def\s{\hbox{s}}
\def\cq{\hbox{c}}
\def\b{\hbox{b}}
\def\t{\hbox{t}}
\def\h{\hbox{h}}
\def\H{\hbox{H}}
\def\A{\hbox{A}}
\def\hpm{\hbox{$\hbox{H}^\pm$}}

\def\Z{\hbox{Z}}

\def\W{\hbox{W}}

\def\mH{\hbox{$m_{\hbox{\rm H}}$}}
\def\mh{\hbox{$m_{\hbox{\rm h}}$}}

\def\mZ{\hbox{$m_{\hbox{\rm Z}}$}}

\def\tb{\hbox{$\tan\beta$}}
\def\sab{\hbox{$\sin^2(\beta-\alpha)$}}
\def\cab{\hbox{$\cos^2(\beta-\alpha)$}}

\def\mtop{\hbox{$m_{\hbox{\rm t}}$}}

\def\eps{\hbox{$\varepsilon$}}

\def\inpb{\hbox{$\hbox{pb}^{-1}$}}

\def\ET{\hbox{$E_T$}}
\def\MAT{\hbox{$M_T$}}
\def\MET{\hbox{$\not{\!\!E}_T$}}

\def\sle{\hbox{$\tilde{\hbox{e}}$}}
\def\ser{\hbox{$\tilde{\hbox{e}}_R$}}

\def\smr{\hbox{$\tilde \mu_R$}}

\def\grav{\hbox{$\tilde{\hbox{G}}$}}

\def\chip{\hbox{$\chi^+$}}
\def\chim{\hbox{$\chi^-$}}

\def\mgu{\hbox{$m_{\tilde g}$}}

\def\mchi{\hbox{$m_\chi$}}

\def\mchap{\hbox{$m_{\chi^+}$}} 
 
\def\msnu{\hbox{$m_{\tilde\nu}$}} 

\def\msq{\hbox{$m_{\tilde q}$}}

\epsfverbosetrue
\title{Moriond '97\\
Electroweak Interactions and Unified Theories\\
Conference Summary}
\author{J.-F. Grivaz\\
{\elevenrm Laboratoire de l'Acc\'el\'erateur Lin\'eaire,}\\ 
{\elevenrm IN2P3-CNRS et Universit\'e de Paris-Sud, F-91405 Orsay}}

\date{}

\begin{document}

\pagenumbering{arabic}
\pagestyle{plain}

\maketitle


\begin{picture}(160,1)
\put(130,85){\parbox[t]{45mm}{\bf LAL 97-60}}
\put(130,80){\parbox[t]{45mm}{August 1997}}
\end{picture}

\vfill

\begin{abstract}
A brief summary of the experimental results presented at this conference is
given.
\end{abstract}

 
\parskip = 3mm plus 1mm

\section{Introduction}

First of all, I would like to apologize for not covering a number of items,
often very interesting, which were discussed during this conference. 
I certainly do not feel competent to address most of the theoretical issues, 
so this talk will be entirely devoted to experimental results. Furthermore, a
number of reviews on specific subjects were presented, which it makes no sense 
to try to summarize: high energy cosmic neutrinos, rare kaon decays, polarized
structure functions. There will be ample opportunities in forthcoming Rencontres
de Moriond to come back to future projects such as B-factories, LHC, long
baseline neutrino experiments, AMS, or to ongoing experiments which are a bit
too young this year to deliver results, such as KTeV, NA48 or the neutron
electric dipole moment measurement at the ILL. These topics will therefore not
be covered either. Even with these restrictions, it will be impossible to do
justice to the vast amount of material which has been presented in the past
week, and I can only reiterate my apologies to those who may feel that their 
contribution is not adequately referred to in the following.

This presentation will be divided in three (unequal) chapters: tests and 
measurements within the standard model, searches and hints beyond the standard
model, and finally neutrino oscillations. Most of the results presented at this
conference were stamped as preliminary; therefore the original contributions 
should be checked in addition to this summary before quoting any results.
For the written version of this talk, the figures have not been incorporated 
since they can be found easily in these proceedings, with the exception of those
belonging to contributions not available to the author at the time of writing.

\section{Within the standard model}

\subsection{Top quark physics}

\centerline{{\it Contributions by D.W. Gerdes}~\cite{gerdes} 
{\it and R. Raja}~\cite{raja}}

Top quarks are produced at the Tevatron in \ppb\ collisions at $\rs=1.8$~TeV,
where both CDF and D0 accumulated $\sim 110$~\inpb. 
With such statistics, the main goals of the experiments are, in this field, the 
measurements of the top quark mass and of its pair production cross section.

Top quarks decay according
to $\t\to\W\b$, so that \ttb\ pair production leads to three final state 
topologies, depending on whether both, one or none of the Ws decay leptonically 
($\W\to\ell\nu$): dileptons, lepton plus jets, all hadronic. Leptons are 
selected as isolated electrons or muons with large transverse energy \ET. The 
presence of neutrinos is inferred from a large amount of missing transverse 
energy \MET. Jets are required to carry substantial \ET, and multijet events
exhibit a spherical pattern. Finally, b-jets are tagged by soft leptons or, in
the case of CDF, by secondary vertices.

In the dilepton topology, the two leptons should not be compatible with a
$\Z\to\lplm$ decay, there should be substantial \MET\ and two additional jets 
should be detected. CDF select nine
events over a background of 2.1, and D0 five over 1.4. Four of the CDF events,
all in the $\e\mu$ channel, have a \MET\ in excess of 100~GeV, which is larger
than the typical expectation from \ttb\ pairs.

In the lepton plus jet topology, at least three jets, \MET\  and a b-tag are 
required.
CDF select 34 events over a background of 9.3, and D0 11 over 2.4. CDF use this
sample for the cross-section measurement, and supplement it with untagged 
four-jet events to reconstruct the top-quark mass (Fig.~2 of \cite{raja}). A 
topological analysis is also performed by D0, requiring at least four jets but 
not imposing any b-tag. The aplanarity and the sum of the jet transverse 
energies are used by means either of cuts to select 19 events over a background 
of 8.7 (Fig.~1 of \cite{gerdes}) from which a cross-section measurement is 
inferred, or of a maximum likelihood fit to extract a measurement of the top 
quark mass (Fig.~3 of \cite{raja}).

The \ttb\ production cross-section is determined to be $7.5^{+1.9}_{-1.7}$~pb 
and $5.5 \pm 1.7$~pb by CDF and by D0, respectively, from the dilepton and 
lepton plus jet samples. For a top mass of 175~\Gcs, the theoretical
expectations are around 5~pb. Using the lepton plus jet topology, CDF and D0 
measure masses of $177 \pm 7$ and $173.3 \pm 8.4$~\Gcs. 

\subsection{Properties of the W boson}

\subsubsection{W mass measurements at the Tevatron}

\centerline{{\it Contributions by A. Gordon}~\cite{gordon} 
{\it and D.Wood}~\cite{wood}}

W bosons are produced via the Drell-Yan process in \ppb\ collisions. The
measurement of the W mass is performed through a fit to the reconstructed 
transverse mass \MAT\ of the $\W\to\ell\nu$ decay. The transverse mass is 
calculated as $M_T^2=(\overrightarrow p_T^\ell + \overrightarrow p_T^\nu)^2
- \vert \overrightarrow U_T \vert^2$, where $\overrightarrow U_T$ is the 
transverse momentum of the recoiling hadronic system. 

These measurements are now limited by systematic errors. The scale of the 
lepton energy is calibrated using events containing $J/\Psi\to\lplm$ decays;
the resolution on the energy of the hadronic system is determined using minimum 
bias events; the model for the transverse momentum distribution of the produced
W bosons is controlled with events containing $\Z\to\lplm$ decays instead.

The CDF results are obtained using $\W\to\mu\nu$ decays (Fig.~5 of 
\cite{gordon}), while D0 use the $\W\to\e\nu$ channel instead (Fig.~5 of 
\cite{wood}). Averaging with the results obtained from Run 1A, CDF measure a 
W mass of $80.375 \pm 0.120$~\Gcs. The D0 result of $80.37 \pm 0.15$~\Gcs\ has 
since then be updated to $80.44 \pm 0.11$~\Gcs, as quoted in \cite{wood}.

\subsubsection{W mass measurements at LEP~2}

\centerline{{\it Contributions by A. Valassi}~\cite{andrea}
{\it and M.A. Thomson}~\cite{thomson}}

There are two very different methods to measure the W mass in \epem\ collisions
at LEP~2. The one 
relies on the behaviour of the W pair production cross section near threshold.
The other explicitly reconstructs the mass of the final state W bosons from
their decay products. Luminosities of about 10~\inpb\ were collected in 1996 by 
each of the LEP experiments both at 161 and 172~GeV.

The measurement at threshold was performed at a centre-of-mass energy of 
161.33~GeV which maximizes the sensitivity of the cross section to the value of
the W mass. (A single measurement at this optimal energy has been shown to be
more efficient than a more detailed scan of the threshold region.) Depending
whether both, one or none of the produced Ws decay leptonically, the final 
state arising from W pair production consists in {\it i)} an acoplanar pair of 
leptons, {\it ii)} an isolated lepton, missing energy and two hadronic jets, or 
{\it iii)} four jets. The first two topologies, which account for 11\% and 44\%
of the final states, respectively, are rather easy to select since they do not 
suffer from any significant standard model background. The four-jet topology, 
on the other hand, is more difficult to disentangle from the large QCD 
background, and multivariate analyses are therefore used to retain sensitivity.
From the cross section measurement of $3.69 \pm 0.45$~pb, averaged over the four
LEP experiments, a W mass value of $80.40 \pm 0.22$~\Gcs\ is inferred.

The cross section measurement was repeated at 172~GeV and the result is well
compatible with the standard model expectation (Fig.~3 of \cite{andrea}). While
the sensitivity to the W mass is reduced, the larger statistics allow a direct
measurement of the W hadronic branching ratio $B_h=(67.0 \pm 2.5)\%$, obtained 
from the comparison of the cross sections in the various topologies. This 
measurement does not compete yet with the indirect determination performed at
the Tevatron using the ratio of the production cross sections for $\W\to\ell\nu$
to $\Z\to\lplm$ ($B_l=(10.43 \pm 0.44)\%$, as reported in \cite{wood}, hence 
$B_h=(68.7 \pm 1.3)\%$). It relies however on fewer theoretical inputs.

The direct reconstruction of the W mass has been performed at 172 GeV where the
statistics is largest. Typically, in the lepton plus two-jet topology where a
neutrino escapes detection, a 2C-fit is performed, imposing equality of the two
W masses; in the four-jet topology, a 5C-fit is performed in a similar fashion,
or a 4C-fit supplemented by a rescaling of the two dijet energies to the
beam energy. There are a number of subtleties such as the choice of jet pairing
in the four-jet topology, the type of functions fitted to the resulting mass
distributions, the bias corrections. Clear mass peaks are observed (Fig.~3 of
\cite{thomson}), and an average W mass of $80.37 \pm 0.19$~\Gcs\ is determined.
This measurement is still limited by statistical errors, but theoretical issues
such as the effect of colour reconnection in the four-jet topology and technical
challenges such as the precise beam energy calibration will become relevant very
soon.

\subsubsection{Summary of top quark and W mass measurements}

The average of the top quark mass measurements at the Tevatron is
$175.6 \pm 5.5$~\Gcs. The Tevatron (plus UA2) average for the W mass is 
$80.41 \pm 0.09$~\Gcs. The impact of these results, which tend to favour 
a light Higgs boson, can be seen in Fig.~6 of \cite{raja}.
The average W mass resulting from the measurements performed at LEP~2 is 
\hbox{$80.38 \pm 0.14$~\Gcs~\cite{thomson}}, well consistent with the value 
from  hadron colliders given above. The grand average is $80.40 \pm 0.08$~\Gcs.

\subsubsection{Triple gauge boson couplings}

\centerline{{\it Contributions by D. Wood}~\cite{wood}
{\it and S. Mele}~\cite{tgc,mele}}

The search for an anomalous $\W\W\gamma$ coupling has been pursued at \ppb\
colliders since many years in final states involving a W boson and a high $p_T$
photon. The results are traditionally expressed in terms of the
$\Delta\kappa_\gamma$ and $\lambda_\gamma$ parameters, which have zero value in
the standard model. Recent D0 results are shown in Fig.~2 of \cite{wood}. They
are perfectly compatible with the standard model and exclude a theory involving
only electromagnetism. With the increased statistics, the search for anomalous
couplings has now been extended to WW, WZ and $\Z\gamma$ production~\cite{wood}.

At LEP~2, W pair production involves, in addition to $t$-channel neutrino
exchange, $s$-channel Z and photon exchange. It is therefore possible to test
the WWZ and $\W\W\gamma$ couplings, but the two are hard to disentangle and thus
a direct comparison with the results from the Tevatron is not easy. Moreover,
there are strong indirect constraints on anomalous couplings resulting from the
precision measurements at LEP~1, except for some specific parameter combinations
called ``blind directions''. The analysis is therefore restricted to such
combinations, {\it e.g.} the $\alpha_{W\phi}$ parameter. Both the total cross
section and the angular distribution of W pair production provide constraints on
the triple gauge boson couplings, as can be seen in Fig.~2 of \cite{tgc}. Here
too, a theory with no WWZ vertex is excluded at more than 95\% CL.

In principle, single W production at LEP~2, through the reaction 
$\epem\to\e\W\nu$ which proceeds dominantly via the $\W\gamma$ fusion mechanism,
could give access to the $\W\W\gamma$ vertex with no contamination from the WWZ
coupling. Such an analysis has been attempted~\cite{mele}, but the results are
still far from competing with those from the Tevatron. 

\subsection{Precision measurements at the Z peak}

\subsubsection{Lineshape, asymmetries and $\sin^2\theta_{eff}$}

\centerline{{\it Contributions by A. B\"ohm}~\cite{boehm}
{\it and  P. Rowson}~\cite{rowson}}

Over four million hadronic events have been collected by each of the four LEP
experiments in the vicinity of the Z peak. Much lower statistics were
accumulated by SLD at the SLC, but with the outstanding specificity of a large
polarization of the electron beam. About 150~k~events were collected with an
average polarization ${\cal P}_e$ of 77\% (to which 50~k with 
${\cal P}_e = 63\%$ from earlier runs can be added).

The combined results from the LEP scan of the Z resonance 
are~\cite{boehm}:\break
\hbox{$m_Z = 91186.3 \pm 1.9$~\Mcs} and
\hbox{$\Gamma_Z = 2494.7 \pm 2.6$~MeV} for the Z mass and width, and 
\hbox{$R_\ell = 20.783 \pm 0.029$}. This last quantity, the ratio of the
hadronic to the leptonic widths, seems a bit high with respect to the standard
model expectation (given \hbox{$\alpha_S = 0.118 \pm 0.003$}). 

The electroweak mixing
angle \sweff\ is determined from a number of independent asymmetry measurements,
the consistency of the results providing a strong test of the standard model. 
As can be seen in Fig.~9 of \cite{boehm}, the $\chi^2$ of the various
determinations is only of 15.1 for 6 DoF, with the LEP \bbb\ asymmetry and 
the left-right polarized asymmetry from SLD contributing most to this large
$\chi^2$ value. The remark can be made that the error on the measurement of the 
left-right asymmetry is dominated by the systematic uncertainty on the beam
polarization. Bringing the SLD measurement of \sweff\ in agreement with the 
LEP average would require a 5\% mismeasurement of the average value of the 
polarization, which seems a bit hard to swallow compared to the
quoted~\cite{rowson} systematic error of less than 1\%. It should also be
remembered that the delicate measurements of the \bbb\ asymmetry and of the 
$\tau$ polarization at LEP are not yet final. With these restrictions in mind,
the grand average~\cite{boehm} is at the moment $\sweff= 0.23151 \pm 0.00022$.

\subsubsection{The story on $R_b$}

\centerline{{\it Contribution by J. Steinberger}~\cite{jack}}

Another controversial precision measurement at the Z peak is that of $R_b$, the
fraction of hadronic Z decays into \bbb. The interest of this quantity is that
it is sensitive to contributions of heavy particles through corrections to the
Zbb vertex (from standard or non standard processes). For instance, the
contribution of the top quark reduces the expected $R_b$ value by 1.2\%.
Last year, the measurement of $R_b$, together with $R_c$, had been said to
exclude the standard model at more than 99\% CL. Such a statement simply 
ignored that systematic errors are often of a highly non-Gaussian nature, and
indeed a lot of effort went, in the past year, into the understanding and the 
control of these systematic errors. 

The most precise measurements of $R_b$ rely on the technique of hemisphere
tagging, in which the lifetime and the mass play the major roles. 
At SLD, the small beam spot characteristic of linear colliders and the 
availability of a vertex detector located at only 3~cm of the beam axis and with
three-dimensional readout allow b purities of 98\% to be achieved with an
efficiency of 35\%, after a simple mass cut as shown in Fig.~2 of \cite{jack}. 
The hemisphere tagging technique allows $R_b$ and the b tagging efficiency to
be determined simultaneously from the data using the total number of 
tagged hemispheres and of events in which both hemispheres are tagged:
$$N_S=2N(R_b\eps_b+R_c\eps_c+R_{uds}\eps_{uds})$$
$$N_D=N(R_b\eps_b^2(1+\rho_b)+R_c\eps_c^2(1+\rho_c)
+R_{uds}\eps_{uds}^2(1+\rho_{uds})).$$
To solve these equations for $R_b$ and $\eps_b$, the hemisphere correlation
$\rho_b$ and the efficiency for charm have to be taken from Monte Carlo and are
responsible for the largest systematic uncertainties. (The uncertainties on the 
other correlations and on $\eps_{uds}$ translate into a very small systematic 
error on $R_b$.) The value of $R_c$ is taken from the standard model, or the
dependence of the result on $R_c$ is explicitly stated. 

Thanks to more detailed assessments of track reconstruction defects, to a
reduction of hemisphere correlations using
techniques such as the reconstruction of separate primary vertices in both
hemispheres, to a more thorough evaluation of physical effects such as gluon
splitting into \bbb, the systematic uncertainties seem to be under a much better
control in the recent measurements than in the earlier ones. Taking only the
most recent ALEPH, DELPHI, OPAL and SLD results leads to ``Jack's
average''~\cite{jack} of \hbox{$R_b = 0.2165 \pm 0.0012$}, in excellent 
agreement with the standard model expectation of $0.2158 \pm 0.0004$. 
This agreement is somewhat spoilt if all existing $R_b$ measurements are 
introduced in the average~\cite{boehm}, \hbox{$R_b = 0.2178 \pm 0.0011$}, but 
the discrepancy with the standard model expectation is now reduced to the 
1.8$\sigma$ level. 

\subsubsection{Results of the global fit}

\centerline{{\it Contribution by A. B\"ohm}~\cite{boehm}}

A global fit to all LEP data~\cite{boehm} leads to an indirect determination of 
the top mass, \hbox{$\mtop = 155 \pm 10$~\Gcs}, in agreement with the direct 
measurement at the Tevatron. The fact that this value is on the low side is 
related to the difficulties with \sweff\ and $R_b$ discussed above. The
tendency, as can be seen in Fig.~10 of \cite{boehm}, is to favour a light Higgs
boson. Taking into account the direct top and W mass measurements from the
Tevatron, a Higgs mass $\mH = 127^{+127}_{-72}$~\Gcs\ is predicted, with 
$\mH < 465$~\Gcs\ at 95\% CL.

\subsubsection{The impact of LEP~2}

\centerline{{\it Contribution by D. Gel\'e}~\cite{gele}}

Cross section measurements in \epem\ collisions were also performed at LEP~2, up
to a centre-of-mass energy of 172~GeV. The agreement with the standard model is
as good as statistics allow, as can be seen in Fig.~9 of \cite{gele}. These
measurements constrain the $\gamma/\Z$ interference term which is normally set 
to its standard model value in the fits to the Z peak data. If this constraint 
is not imposed, the precision on the Z mass is only 6.1~\Mcs\ from LEP~1 data, 
and becomes 3.1~\Mcs\ using the LEP~2 (and TOPAZ) data in addition~\cite{gele}. 
This is not very far from the 1.9~\Mcs\ precision achieved when setting the 
$\gamma/Z$ interference term to its standard model value.

\subsection{Highlights in $\tau$ and b physics}

Here, only a few highlights in $\tau$ and b physics will be sketched, for
completeness.

A factor of four improvement in the precision on Michel parameters in $\tau$
decays has been achieved by CLEO~\cite{jessop}. 

Lepton universality is now tested at the 0.3\% level in $\tau$
decays~\cite{gentile}. The comparison of the branching ratios for 
$\tau\to\e\nunu$ and $\tau\to\mu\nunu$ provides a test of e$\mu$ universality 
at that level. The compatibility of the $\tau$ decay leptonic branching ratio, 
$(17.80 \pm 0.05)\%$ for a massless fermion, with the $\tau$ lifetime, 
$290.5 \pm 1.2$~fs, provides a test at the same level, given the value of the 
$\tau$ mass measured at BES.

The vector and axial-vector structure functions in $\tau$ hadronic decays have 
been measured separately, allowing an improvement of 30\% on the theoretical 
error on \hbox{$g_\mu-2$~\cite{alemany}}, which is of interest in view of the
forthcoming measurement of that quantity at Brookhaven.

A collection of rare B decays, mediated by penguin diagrams, has been 
investigated by CLEO~\cite{ogrady}. Mostly, limits have been set, but the 
process $\hbox{B}^+\to\eta'\hbox{K}^+$ was observed, with a branching ratio of 
\hbox{$(7.8^{+2.7}_{-2.2} \pm 1.0) 10^{-5}$}. Surprisingly enough, a measurement
of $b\to\s\gamma$ was performed at LEP by ALEPH, 
\hbox{$(3.4\pm 0.7\pm 0.9)10^{-4}$}~\cite{parodi}. 
The value of $\vert V_{cb} \vert$ is determined to be 
$0.0368 \pm 0.0022 \pm 0.0012$, using the 
$\hbox{B}\to\hbox{D}^\ast/\hbox{D}\ell\nu$ decays~\cite{parodi}.

All exclusive b hadron lifetimes are measured, and their ratios are found to be
compatible with expectation, except for the $\Lambda_b$ lifetime which remains
low. A new measurement of the $\hbox{B}^0$ lifetime was performed with an 
accuracy of 56~fs by DELPHI~\cite{parodi}, using the signature of the slow pion 
from $\hbox{D}^\ast\to\hbox{D}\pi$ in the decay 
$\hbox{B}^0\to\hbox{D}^\ast X\ell\nu$. A similar precision was reached by
CDF~\cite{troconiz}.

B-mixing has been studied at LEP~\cite{parodi}, SLD~\cite{usher} and
CDF~\cite{troconiz}. A variety of methods is used to measure $\Delta m_d$,
leading to the LEP average of \hbox{$0.463 \pm 0.018$~ps$^{-1}$}. (The breakdown
of systematic errors, necessary for a proper averaging, was not available from 
CDF and SLD at the time of the conference.) The ALEPH and DELPHI combined lower
limit on $\Delta m_s$ is 9.2~ps$^{-1}$~\cite{parodi}, which becomes interesting 
not only from a technical point of view.

\section{Beyond the standard model}

\subsection{Supersymmetry}

\subsubsection{The standard model Higgs boson}

\centerline{{\it Contributions by P. Gay}~\cite{gay}
{\it and S. Rosier-Lees}~\cite{sylvie}}

The search for the ``Standard Model Higgs Boson'' really belongs to this section
on Supersymmetry. This is because, in the minimal standard model, there is
essentially no room, given the large top quark mass, for a Higgs boson light 
enough to be discovered at LEP~2, the only place where this search can be
conducted efficiently these days. Moreover, in large regions of the parameter 
space of the MSSM (the minimal supersymmetric extension of the standard model), 
the properties of the lightest Higgs boson make it indistinguishable in practice
from its standard model equivalent.

The main production mechanism is the Higgsstrahlung process, $\epem\to\H\Z$,
which leads to various topologies, depending on the H and Z decay modes, of
which three are the most important. Acoplanar jets result from the $\Z\to\nunu$
and $\H\to\hbox{hadrons}$ decays; the $\Z\to\lplm$ decay leads to two isolated 
energetic leptons in a hadronic environment instead; a four-jet topology is
reached when both the Higgs and the Z decay into hadrons. A crucial feature
affecting the searches in the various topologies is the large decay branching
ratio of the Higgs into \bbb, 85\%. 

Although most of the signal ends up in a four-jet final state, this topology 
had not been considered at LEP~1 because of the overwhelming background from 
hadronic Z decays, with a typical signal to background ratio of $10^{-6}$. 
At LEP~2 on the contrary, this ratio is of order $10^{-2}$, which renders
worthwhile the search in this channel. An efficient b-tagging is the key to
the reduction of both the QCD and the WW backgrounds. This tool is also
instrumental in the acoplanar jet topology to eliminate the backgrounds from W
pairs (with one W decaying into hadrons, the other into $\tau\nu$), and from
single W production in the reaction $\epem\to\e\W\nu$, where the spectator
electron remains undetected in the beam pipe. In all channels, the constraint
that the decay products of the Z should have a mass compatible with \mZ\ is 
also highly discriminating, a feature which could not be used at LEP~1 where 
the final state Z was produced off-shell.

For a 70~\Gcs\ Higgs boson mass, about 10 events would have been produced in
each of the experiments. Typically, efficiencies of $\sim 30\%$ are achieved
for a background expectation of one event. No signal was observed, resulting 
in the case of ALEPH in a mass lower limit of 70.7~\Gcs, as shown in Fig.~1a 
of \cite{gay}. When the results of the four experiments are combined, a 
sensitivity in excess of 75~\Gcs\ should be reached.

\subsubsection{Supersymmetric Higgs bosons}

\centerline{{\it Contributions by P. Gay}~\cite{gay}
{\it and S. Rosier-Lees}~\cite{sylvie}}

In the MSSM, two Higgs doublets are needed, which leads to three neutral Higgs
bosons, the CP-even h and H, and the CP-odd A, and to a pair of charged Higgs 
bosons \hpm. While H and \hpm\ are expected to be out of the LEP~2 reach, h 
should be fairly light. In addition to \mh, the ratio \tb\ of the two Higgs 
field vacuum expectation values is the important parameter for phenomenology.
Compared to the standard model case, the Higgsstrahlung production cross 
section is reduced by the factor \sab, where $\alpha$ is the mixing angle in 
the CP-even Higgs sector. The results from the standard model Higgs searches 
reported above can therefore be turned into limits on \mh\ as a function of 
\sab. These limits are most constraining for low values of \tb\ ({\it i.e.} for
\tb\ close to unity), as can be seen in Fig.~1b of \cite{gay}.

For large values of \tb, the complementary process $\epem\to\h\A$ becomes
dominant, with a cross section proportional to \cab, in which case h and A 
are almost mass degenerate. Since both h and A decay predominantly into \bbb,
the main final state consists in four b quark jets. The $\toto\bbb$ 
topology has also been addressed, adding some sensitivity to the search.
The result obtained by ALEPH combining the searches for both the hZ and hA 
final state is shown in Fig.~1b of \cite{gay}. It can be seen that, although 
the boundaries of the region theoretically allowed depend on additional 
parameters of the model, in particular on the mixing in the top squark sector,
the experimental result shows hardly any dependence on those. A lower mass 
limit of 62.5~\Gcs\ holds for both h and A, for any $\tb>1$. 

\subsubsection{Supersymmetric particles: the standard scenario}

\centerline{{\it Contribution by S. Rosier-Lees}~\cite{sylvie}}

In the standard scenario, R-parity is conserved and the lightest supersymmetric
particle (LSP) is a neutralino, $\chi$. A variety of searches for supersymmetric
particles have been performed at LEP~2, as reported in detail in \cite{sylvie}.
Here the example of chargino pair production, $\epem\to\chip\chim$, will be 
sketched for illustration.

In most of the parameter space, the decay modes of charginos in the mass range 
relevant at LEP~2 are $\chip\to\ell^+\nu\chi$ and 
$\chip\to\hbox{q}\,\overline{\hbox{q}}\,'\chi$. The topologies arising from pair
production are therefore {\it i)} an acoplanar lepton pair, {\it ii)} an
isolated lepton in a hadronic environment or {\it iii)} multijets. In all cases,
there should be substantial missing energy. The analyses addressing these
various final states are further split according to the \chip--$\chi$ mass
difference: for small mass differences, the main background comes from \gaga\
interactions, while for very large mass differences, the signal resembles W pair
production. No signal was detected above background in any of those searches. 

Since the production cross section and the decay branching ratios are quite
model dependent, it is difficult to derive a hard limit for the chargino mass.
For gaugino-like charginos, the $\chi$ mass is about half the chargino mass, and
the cross section is largest if sneutrinos are heavy; in that case, the
kinematic limit of 86~\Gcs\ is reached. The limit is lowered to 72~\Gcs,
allowing for any $m_{\tilde\nu}>\mchap$. For higgsino-like charginos, the mass
difference tends to be small but the cross section does not depend on 
$m_{\tilde\nu}$; in that case, the mass limit is 80~\Gcs\ for mass differences 
in excess of 5~\Gcs.

\begin{figure}[htbp]
\begin{center}
\epsfig{file=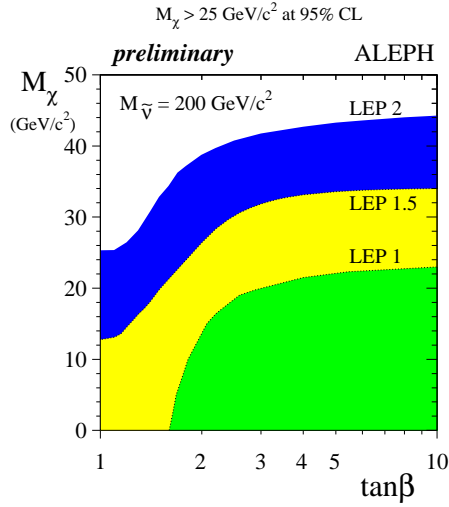,
height=8.cm,bbllx=0pt,bblly=25pt,bburx=500pt,bbury=650pt}
\end{center}
\caption{\label{fig:mchi}Lower limit on the mass of the lightest neutralino as a
function of \tb, valid for $\msnu>200$~\Gcs.}
\end{figure}

In the MSSM, the results on chargino searches at LEP~2 can be combined with 
the LEP~1 constraints on neutralinos to set limits on the mass of the lightest 
neutralino. Assuming heavy sneutrinos, a lower limit of 25~\Gcs\ is obtained,
irrespective of \tb. The limit increases with \tb\ as shown in
Fig.~\ref{fig:mchi}. 

\begin{figure}[htbp]
\begin{center}
\epsfig{file=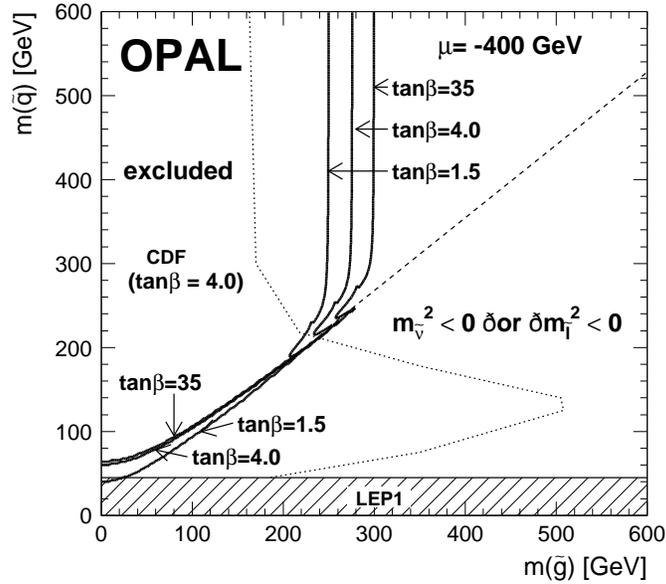,
height=8.cm,bbllx=0pt,bblly=380pt,bburx=415pt,bbury=750pt}
\end{center}
\caption{\label{fig:opal}Indirect limits in the (\msq,\mgu) plane, derived
from the chargino searches at LEP~2 for $\mu=-400$~\Gcs\ and for various values
of \tb. The CDF limit for $\tb=4$ is also shown.}
\end{figure}

Assuming unification of all gaugino masses and of all squark and slepton masses
at the GUT scale, the constraints from LEP~2 can be compared with those inferred
at the Tevatron from the absence of any squark or gluino signal. This has been
done by OPAL, as shown in 
Fig.~\ref{fig:opal}.
It can be seen that,
for large squark masses, the indirect LEP~2 limit on the gluino mass is of 
almost 280~\Gcs, well above the direct limit of 160~\Gcs. (The exact values 
depend on \tb\ and $\mu$; the CDF parameter choice has been used for 
comparison).

Other LEP~2 mass limits are 67~\Gcs\ for \smr\ and 75~\Gcs\ for \ser\ (assuming
\tb=2), both for $\mchi=35$~\Gcs; the stop mass limit varies between 69 and
75~\Gcs, depending on \mchi\ and on the mixing angle in the stop sector.

\subsubsection{Photonic signals of supersymmetry}

\centerline{{\it Contributions by P. Azzi}~\cite{azzi} 
{\it , J.D. Hobbs}~\cite{hobbs} {\it and J.H. Dann}~\cite{dann}}

In 1995, CDF reported the observation of a spectacular event containing two 
electrons and two photons, all with \ET\ in excess of 30~GeV, and with a missing
transverse energy of 53~GeV. There is no convincing explanation of this event
within the standard model, which triggered a lot of theoretical activity.
Two classes of supersymmetric interpretation were proposed, both advocating that
this event originates from selectron pair production. In the light gravitino 
scenario, the usual $\sle\to\e\chi$ decay takes place, followed by 
$\chi\to\gamma\grav$. In the almost standard MSSM scenario, the $\sle\to\e\chi'$
decay takes place instead, with $\chi'\to\chi\gamma$. Here, the GUT relation
among gaugino masses has to be dropped, and $\chi'$ is an almost pure photino
while $\chi$ is an almost pure higgsino. If one of these interpretations is
correct, there should be many other channels leading to final states containing 
two photons and missing \ET. This signature has been searched inclusively by 
both CDF and D0, and no signal was observed (other than the previously reported 
event). This can be seen In Fig.~2 of \cite{azzi} and in Fig.~1 of \cite{hobbs}.
However, the sensitivity of these searches is not sufficient to completely rule 
out either of the two proposed scenarii.

\begin{figure}[htbp]
\begin{center}
\epsfig{file=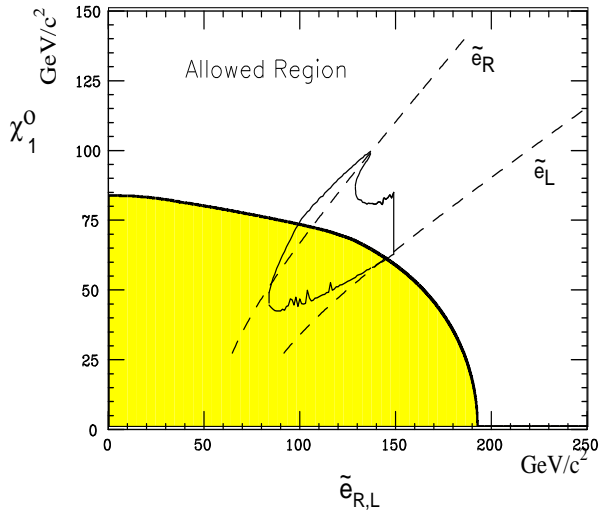,
height=8.cm,bbllx=150pt,bblly=220pt,bburx=470pt,bbury=500pt}
\end{center}
\caption{\label{fig:dann}Region in the ($m_{\tilde e}$,\mchi) plane excluded 
by the ALEPH search for acoplanar photons from the reaction
$\epem\to\chi\chi\to\gamma\gamma\grav\grav$, for a bino neutralino. Also shown
is the region kinematically compatible with the CDF event.}
\end{figure}

The usual phenomenology of supersymmetry at LEP is also deeply modified in these
two models, the clearest signature becoming a pair of acoplanar photons with
missing energy. In the light gravitino scenario, this final state results from
$\epem\to\chi\chi$, with $\chi\to\gamma\grav$, while in the higgsino LSP
scenario, it is reached through $\epem\to\chi'\chi'$, followed by
$\chi'\to\chi\gamma$. Both reactions proceed dominantly via t-channel selectron
exchange. The signatures are somewhat different in the two scenarii because the
light gravitino is practically massless, in contrast to the higgsino LSP.
Requiring two energetic photons at large angle with respect to the beam 
eliminates the standard model background from $\epem\to\gamma\gamma\nunu$. 
The absence of any signal allows $\chi$ mass limits in excess of 70~\Gcs\ to be
set in the light gravitino scenario, for selectron masses around 100~\Gcs. This
excludes half of the domain compatible with the kinematics of the CDF event, as
can be seen in 
Fig.~\ref{fig:dann}.
The constraints on the higgsino LSP
scenario are much milder. Single photons could also arise from the processes
$\epem\to\chi\grav$ or $\epem\to\chi'\chi$. No significant effect was observed
beyond the standard model expectation from $\epem\to\gamma\nunu$~\cite{wilson}.

\subsection{Four-jet events at LEP}

\centerline{{\it Contribution by G.W. Wilson}~\cite{wilson}}

Last year at Moriond, the ALEPH collaboration reported the observation of an
excess of four-jet events in the data collected at 130--136~GeV. Choosing the
pairing of jets such that the dijet mass difference is smallest, nine events
were found to cluster close to 105~\Gcs\ in the dijet mass sum, while less than
one were expected. The other collaborations did not observe any similar effect.

A working group was set up by the LEPC, involving members from the four
collaborations, in order to study if this discrepancy could be explained by some
experimental artefact. The conclusion is twofold: the ALEPH events are real and
do cluster in the dijet mass sum as reported; the other collaborations have similar
sensitivity to such events and would have seen them if they had been present in 
their data samples.

With the additional data collected at 161 and 172~GeV, the effect appears 
enhanced in the ALEPH data and still does not show up elsewhere, as can be seen
in Fig.~3 of~\cite{wilson}. In the mass window from 102 to 110~\Gcs, ALEPH find
altogether 18 events for a background expectation of 3.1, while DELPHI, L3 and 
OPAL together find nine events for an expectation of 9.2.

The probability that the ALEPH observation arises from a fluctuation of the 
standard model is extremely low, and so is the probability that the three other
collaborations see nothing if the ALEPH signal is real (precise numbers are 
of little interest at this point). Only additional data will allow this issue
to be settled unambiguously.

\subsection{High $Q^2$ events at HERA}

\subsubsection{What H1 and ZEUS see}

\centerline{{\it Contributions by E. Perez}~\cite{perez}
{\it and B. Straub}~\cite{straub}}

At HERA, 27.5~GeV positrons collide on 820~GeV protons, the centre-of-mass 
energy thus being $\rs=300$~GeV. The following results are based on 14 and 
20~\inpb of data collected by the H1 and ZEUS experiments, respectively.

The kinematics of deep inelastic electron-proton scattering is sketched in 
Fig.~1 of \cite{perez}. The relevant variables are $Q^2$, $x$ and $y$, the three
being related by $Q^2=xys$. The centre-of-mass energy in the electron-quark
collision is related to $x$ by $m_{eq}^2=xs$, and the scattering angle in that
frame is such that $y=(1+\cos\theta_e^\ast)/2$. 

Four independent quantities can be measured in the laboratory: the electron
energy and angle with respect to the beam, $E_e$ and $\cos\theta_e$; the same
variables for the hadronic jet, $E_j$ and $\cos\theta_j$. Two of these
quantities are sufficient to reconstruct the kinematics of the event. H1 choose
the electron variables $E_e$ and $\cos\theta_e$, while ZEUS rather use the two 
angles, $\theta_e$ and $\theta_j$. Both methods have their virtues and defects. 
The electron energy is sensitive to the absolute calibration of the calorimeter;
the measurement using the angles is more affected by initial state radiation.
Both collaborations therefore use an alternative method as a check. They claim
that their systematic errors are controlled at a level of 8.5\% at high $Q^2$
and that the backgrounds are negligible in that regime. 

\begin{figure}[htbp]
\begin{center}
\epsfig{file=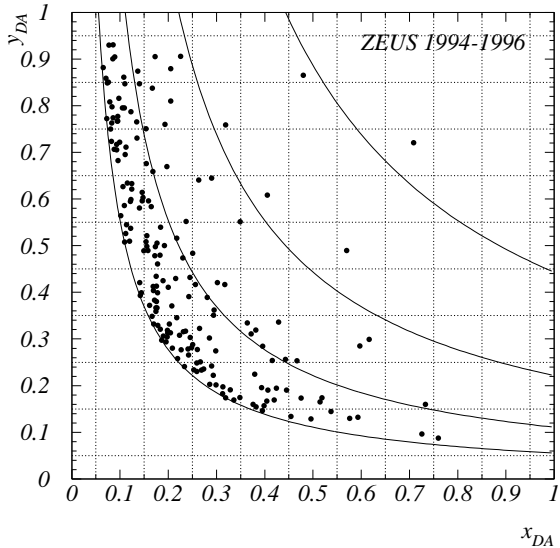,
height=8.cm,bbllx=0pt,bblly=0pt,bburx=550pt,bbury=550pt}
\end{center}
\caption{\label{fig:zeus}Distribution of the events in the ($x$,$y$) plane. The
lines correspond to constant $Q^2$ values of 5000, 10000, 20000 and 
40000~GeV$^2$.}
\end{figure}

The observation of ZEUS is summarized in 
Fig.~\ref{fig:zeus}.
The number of events observed for $Q^2>5000$~GeV$^2$ is 191, for an expectation 
of 196. However, for $Q^2>35000$~GeV$^2$, only 0.145 events are expected while
two are observed; and four events are found with $x>0.55$ and $y>0.25$, to be
compared to an expectation of 0.91. 

Similar information from H1 is available in Fig.~3 of \cite{perez}, where the 
$m_{eq}$ variable is used rather than $x$. Specific projections with additional
cuts are displayed in Fig.~5 of \cite{perez}, from which it can be inferred that
an excess compared to the standard model expectation is observed for
$Q^2>15000$~GeV$^2$. This excess is more apparent for large values of $m_{eq}$
and of $y$, and the effect is maximal for $m_{eq}$ in a 25~\Gcs\ window around 
200~\Gcs\ and for $y>0.4$ where seven events are observed while only one is
expected.

Even if it seems clear that both experiments find more events at high $Q^2$ than
they had expected, the question of the compatibility of their observations can 
be raised. Indeed, the largest effects seen by the two experiments take place in
disconnected regions. In the high $x$ -- high $y$ corner where ZEUS count four
events, H1 find none. In the mass window where H1 count seven events, ZEUS
find two. The obvious conclusion in this somewhat confused situation is: ``Wait
(for more data) and see...''

\eject
\subsubsection{Related investigations}

\centerline{{\it Contributions by J.H. Hobbs}~\cite{hobbs}
{\it and S. Rosier-Lees}~\cite{sylvie}}

If the accumulation of the H1 high $Q^2$ events in a narrow electron-quark mass 
window is not due to a statistical fluctuation, it could be interpreted as a 
signal of the resonant production of a first generation leptoquark or of a 
squark with R-parity violation. In the latter case, the coupling $\lambda'$
at the ed$\tilde{\hbox{q}}_u$ vertex would be of the order of 
0.03/$B(\tilde{\hbox{q}}\to\e^+\d)$. With such a large coupling value, 
the limits on neutrinoless double beta decay exclude a $\tilde{\u}$ 
interpretation, and those on the $\hbox{K}^+\to\pi^+\nunu$ decay almost rule out
a $\tilde{\cq}$, leaving a $\tilde{\t}$ as sole candidate.

Such a squark or leptoquark could be pair produced in \ppb\ collisions, with a
cross section independent of the value of the $\lambda'$ coupling. There are 
three possible final states to consider, depending on whether both, one or none 
of the two squarks/leptoquarks decay into eq, the alternative decay mode being 
$\nu$q. Thus, the final state may consist of two high \ET\ electrons and 
two jets, of one high \ET\ electron, \MET\ and two jets, or of two jets and 
large \MET. Only D0 reported on a search for first generation leptoquarks, the 
result of which is shown in Fig.~4 of \cite{hobbs}. This search has no 
sensitivity to masses as high as 200~\Gcs, but it was not optimized for that 
mass range and progress is to be expected soon.

\eject
The same interpretation of the high $Q^2$ events in terms of squarks or
leptoquarks leads to the prediction that the cross section for $\epem\to\qqb$
should be distorted by a contribution from the $t$-channel exchange of such an 
object. This has been investigated by OPAL at LEP~2~\cite{sylvie}, but the 
present sensitivity is an order of magnitude larger than what would be needed. 

\section{Neutrino oscillations}

Before summarizing the neutrino oscillation results presented at this
conference, it may be worth recalling a few basics. In a two-flavour oscillation
scheme, the probability for a neutrino born as $\nu_a$ to be detected as $\nu_b$
reads
$${\cal P}_{ab}=\sin^2 2\theta_{ab} \sin^2 (1.27\frac{L}{E}\Delta m^2_{ab}),$$
where $L$ is the distance of the detector to the source and $E$ is the neutrino 
energy, in m/MeV or km/GeV; $\Delta m^2$ is measured in eV$^2$. 
If a given experiment reaches a sensitivity ${\cal P}$ for the oscillation 
probability, this translates into a limit of $2{\cal P}$ on 
$\sin^2 2\theta$ at large $\Delta m^2$, and of 
$\sqrt{{\cal P}}/(1.27L/E)$ on $\Delta m^2$ for $\sin^2 2\theta = 1$. 
The reach in $\Delta m^2$ is therefore characterized by the value of $L/E$. As
can be seen in Table~\ref{tab:nuosc}, the various experiments cover a huge
range of $L/E$ values, with only little overlap, a feature which renders cross
checks difficult. 

\begin{table}[htbp]
\begin{center}
\begin{tabular}{|c|c|l|c|}
\hline
Experiments & $L/E$ in m/MeV & $L/E$ & Oscillation type \\  
\hline
Solar neutrinos & $10^{11}/1$ & $10^{11}$ & $\nu_e\to X$ \\
Reactors & $50/5$ &  $10^1$ &  $\bar\nu_e\to X$ \\ 
Atmospheric neutrinos & $10^{4\div 7}/10^3$ & $10^{1\div 4}$ & $\nu_\mu\to X$ \\
Beam stops & $50/50$ & $10^0$ & $\bar\nu_\mu\to\bar\nu_e$ \\
CHORUS -- NOMAD & $10^3/10^4$ & $10^{-1}$ & $\nu_\mu\to\nu_\tau$ \\
Long base line~\cite{bouchez} & $10^6/10^4$ & $10^2$ & 
                    $\nu_\mu\to(\nu_e$ or $\nu_\tau$ or $X)$ \\
\hline
\end{tabular}
\end{center}
\caption{Typical $L/E$ values for the various kinds of neutrino 
oscillation experiments. Also indicated are the relevant types of oscillation.}
\label{tab:nuosc}
\end{table}

There are at the moment three independent indications for neutrino
oscillations:

\begin{itemize}
\item Solar neutrinos, with $\Delta m^2 \sim 10^{-5}$ and $\sin^2 2\theta \sim
10^{-2}$ in the MSW interpretation;
\item Atmospheric neutrinos, with $\Delta m^2 \sim 10^{-2}$ and $\sin^2 2\theta
\sim 1$;
\item LSND, with $\Delta m^2 \sim 1$  and $\sin^2 2\theta \sim 10^{-2}$.
\end{itemize}

\subsection{Solar neutrinos}

\centerline{{\it Contributions by C. Galbiati}~\cite{galbiati}
{\it and Y. Takeushi}~\cite{takeushi}}

The radiochemical experiments using Gallium are the only ones with a threshold 
low enough to be sensitive to the pp neutrinos. The final results from Gallex 
were reported~\cite{galbiati}. The measurement is $69.7 \pm 6.7 \pm 4.2$ solar 
neutrino units, to be compared with expectations in the $110\div 140$ range 
from the solar models.

First results from Super-Kamiokande, based on 102 days of data taking were
presented~\cite{takeushi}. Super-Kamiokande is a huge water 
$\check{\hbox{C}}$erenkov
detector with a fiducial mass of 22,000 tons, sensitive only to the Boron
neutrinos. The early observation by Kamiokande of neutrinos coming from the 
direction of the sun is beautifully confirmed, as can be seen in Fig.~2 of 
\cite{takeushi}. The deficit by a factor 0.44 with respect to expectation 
remains and, as shown in Fig.~4 of \cite{takeushi}, there is no conspicuous 
energy modulation. No results on atmospheric neutrinos were reported.

Taking these results together with those from the Homestake Chlorine experiment,
there is essentially no room left for the Beryllium neutrinos, which may be
accommodated by the MSW effect. 
Crucial tests will be provided by Borexino, which
is aimed at the real time detection of Beryllium neutrinos, by Super-Kamiokande
when the statistics are sufficient to allow fine distortions of the energy 
spectrum to be detected, and by Sudbury which should be able to measure the 
neutral to charged current ratio.

\subsection{Beam stops}

\centerline{{\it Contributions by K. Eitel}~\cite{eitel}
{\it and D.H. White}~\cite{white}}

The principle of a neutrino oscillation experiment at a beam stop is very 
simple. A high intensity proton beam is absorbed in a target. The positive
pions produced come to rest and decay into $\mu^+\nu_\mu$ while the negative 
ones are absorbed by nuclear capture. The decay muons also come to rest and 
decay into $\e^+\nu_e\bar\nu_\mu$. 
The only particles reaching the detector, located at a
few tens of metres from the target, are the neutrinos, of which all species are
present except for $\bar\nu_e$.

Both KARMEN and LSND search for the appearance of this forbidden neutrino 
which could originate from a $\bar\nu_\mu\to\bar\nu_e$ oscillation. The reaction
used for detection is $\bar\nu_e\hbox{p}\to\e^+\hbox{n}$. The positron should be
detected with an energy limited to $\sim 50$~MeV. The signature of the neutron 
is a delayed 2~MeV $\gamma$-ray from the neutron capture. The advantage of
KARMEN is the sharp time structure of the proton beam. LSND benefits from a
larger detector mass and from particle identification.

The by now well known signal of oscillation observed by LSND is shown in Fig.~2
of \cite{white}. It corresponds to an oscillation probability of 
$4\cdot 10^{-3}$.
The new information reported in \cite{white} is that a very preliminary 
3$\sigma$ signal of 11 events over a background of 11 is also observed in the 
in-flight decays. These come from an imperfect containment of the produced 
positive pions in the water target. Here, a $\nu_\mu\to\nu_e$ oscillation 
signal is searched for, with very different signatures (harder electron energy 
spectrum, no neutron) and systematics. 

It will be interesting to see what happens to this new signal; and the
results from the upgraded KARMEN experiment~\cite{eitel}, where the veto 
against cosmic muons has been greatly improved, are also eagerly awaited.

\subsection{CHORUS and NOMAD}

\centerline{{\it Contributions by M. Vander Donckt}~\cite{muriel}
{\it , B.A. Popov}~\cite{popov} {\it and A. de Santo}~\cite{desanto}}

The two CERN neutrino experiments, CHORUS and NOMAD, are aimed at the detection
of $\nu_\mu\to\nu_\tau$ oscillations in the cosmologically relevant mass domain
of a few eV. This mass range can also be expected, given the indications from
the solar neutrino experiments and invoking the see-saw mechanism.

The principle of CHORUS~\cite{muriel} is the direct observation of the $\tau$
decay vertex, hence the use of the emulsion technique. Considerable R\&D efforts
went into the automation of the scanning of those emulsions. A limit of 
$8\cdot 10^{-3}$ 
on $\sin^2 2\theta$ has been achieved for large $\Delta m^2$, based 
on the pilot analysis of a small fraction of the data and using the 
$\tau\to\mu$ channel only.
In the case of NOMAD~\cite{popov}, the goal is to fully reconstruct the event
kinematics to identify the presence of a $\tau$ by the missing momentum carried 
away by its decay neutrino(s), hence the need for a low density target and for
excellent momentum and energy resolutions. The limit on $\sin^2 2\theta$
presently achieved is $4\cdot 10^{-3}$, 
at the level of the best previous result. 

When they have collected and analysed their full statistics, both experiments
should reach the $2\cdot 10^{-4}$ level. Projects to increase the sensitivity
by an order of magnitude are submitted both at Fermilab and at CERN, as 
reviewed in \cite{bouchez}.

A search for $\nu_\mu\to\nu_e$ oscillations was also performed by
NOMAD~\cite{desanto}. In contrast to the $\nu_\mu\to\nu_\tau$ case, this search
is not background free since it suffers from the $\nu_e$ contamination in the
$\nu_\mu$ beam. The analysis therefore relies on a precise knowledge of the
$\nu_e$ component of the beam. Detailed Monte Carlo calculations of the CERN 
neutrino beam are available, but the $\nu_e$ flux can also be determined 
from the data itself by a reconstruction of the various sources of electron 
neutrinos. This is shown in Fig.~2 of \cite{desanto}: Using the appropriate 
charged current interactions, the dominant $\hbox{K}^+$ component 
can be monitored using the high energy tail of the $\nu_\mu$ spectrum; 
similarly, the $\hbox{K}^0_L$ component can be derived from the
$\bar\nu_e$ spectrum. 

It can be seen in Fig.~4 of \cite{desanto} that the set of $\sin^2 2\theta$ 
and $\Delta m^2$ values giving the best fit to the LSND data is clearly 
excluded. 
Altogether, as shown in Fig.~5 of \cite{desanto}, this analysis excludes 
the large $\Delta m^2$ domain allowed by LSND. Accessing the lower $\Delta m^2$ 
region would need a medium baseline experiment, as discussed in \cite{bouchez}.

\section{Conclusion}

Is it useful to conclude a summary ? It can certainly be said that this has
been one of the very good Rencontres de Moriond. Snow was excellent, and the
weather superb (too bad for the summary speaker). Many enthusiastic young --- 
and sometimes older --- speakers were given an opportunity to defend their 
work in front of a demanding audience. Thanks to you all, a lot of fresh
high quality results were presented. Last but not least, Tran's hospitality has
been, as usual, perfect. We all look forward to Moriond '98.

\end{document}